\documentclass[prd,aps,showpacs,twocolumn,floatfix,nofootinbib]{revtex4}
\usepackage{amssymb,graphicx}
\usepackage{amsmath}
\usepackage{epsfig}
\usepackage[usenames]{color}
\usepackage{multirow}

\newcommand{\beqn}{\begin{eqnarray}}
\newcommand{\eeqn}{\end{eqnarray}}
\newcommand{\beq}{\begin{equation}}
\newcommand{\eeq}{\end{equation}}

\def\s1pr{\mathcal{I}^{1+}_{\rm R}}

\def\t{\tilde}

\newcommand{\w}[1]{\boldsymbol{#1}}

\newcommand{\Liec}[1]{{\mathcal{L}}_{\w{#1}}\,}

\newcommand{\gm}{\gamma}

\newcommand{\tD}{{\tilde D}}

\newcommand{\hA}{{\hat A}}

\begin{document}

\title{Conformal thin-sandwich solver for generic initial data}

\author{William E.\ East, Fethi M.\ Ramazano\u{g}lu, and Frans Pretorius}
\affiliation{
Department of Physics, Princeton University, Princeton, NJ 08544, USA \\
}
\begin{abstract}
We present a new scheme for constructing initial data for the Einstein
field equations using the conformal thin-sandwich formulation that does not assume 
conformal flatness or approximate Killing vectors. 
This includes a method for determining free data
based on superposition, as well as a way to handle black hole singularities without 
excision.  We numerically solve the constraint equations using a multigrid algorithm 
with mesh refinement.  We demonstrate the efficacy of the method with
initial data solutions for several applications: a quasicircular binary black hole merger,
a dynamical capture black hole-neutron star merger, and an ultrarelativistic collision.  
\end{abstract}

\pacs{04.20.Ex,04.25.D-,04.25.dg,04.25.dk,04.30.Db}

\maketitle

\section{Introduction}
\label{intro}
The purview of numerical relativity has extended to include not only relativity theory, but also a wide range of
other topics.  Motivated by current and upcoming efforts to detect gravitational waves~\cite{LIGO,VIRGO,GEO,Somiya:2011me,Sathyaprakash:2012jk},
there has been extensive work on mergers of binary compact objects (BCOs)~\cite{co_binaries_review} including 
binary black holes
(BH-BH)~\cite{BBHreview,Hannam:2009rd,Centrella:2010mx,bbh_review_2011}, binary neutron stars~\cite{lrr-2012-8},
and black hole-neutron star (BH-NS) systems~\cite{BHNSreview,Duez:2009yz}.
In addition to binaries in quasicircular orbits, there have also been studies of eccentric 
binaries~\cite{PhysRevD.82.024033,PhysRevLett.102.041101,Ecc1,bhns_astro_paper,Ecc2,PhysRevD.78.064069,
Pretorius:2007jn,PhysRevLett.101.061102,PhysRevD.77.081502,Gold:2011df,Gold:2012tk} 
as may arise, for example, 
from dynamical capture.
Other works of interest to astrophysics include gravitational collapse of stars~\cite{Ott:2008wt,lrr-2011-1},
black hole accretion~\cite{lrr-2008-7}, and the nature of cosmological 
singularities~\cite{lrr-2002-1,Garrison:2012ex}.
Aside from astrophysical systems, numerical
relativity has also emerged as a useful tool to explore various concepts
in gravity and high energy physics~\cite{2012arXiv1201.5118C}, such as critical collapse~\cite{crit_collapse_review}, 
ultrarelativistic collisions~\cite{Sperhake:2008ga,Shibata:2008rq,Choptuik:2009ww,Okawa:2011fv,Witek:2010az}, 
the gauge/gravity duality~\cite{Chesler:2008hg,Chesler:2010bi,Chesler:2011ds,Bantilan:2012vu,Buchel:2012gw}, 
gravity and black holes in higher dimensions~\cite{Wiseman:2011by,Shibata:2009ad,Lehner:2010pn}, and 
the (in)stability of anti-de-Sitter spacetime~\cite{Bizon:2011gg}.
In all these applications, a necessary ingredient 
is a good method for constructing initial data (ID).  
Here we present a new initial data solver, based on the conformal thin-sandwich (CTS)~\cite{YorkCTS}
formulation, which we have designed 
to be more generally applicable to a range of physical 
scenarios by avoiding symmetry or simplifying assumptions.

There has been extensive research on the problem of constructing ID for general relativity, and 
detailed reviews can be found in~\cite{1989fnr..book...89Y,lrr-2000-5,2004gr.qc....12002P,gourgoulhon}. 
Early attempts at solving the initial data problem relied on certain assumptions to
make the mathematical formulation of the problem more tractable, such as conformal flatness and maximal
slicing.
The widely used Bowen-York solution~\cite{BowenYork} is one
such example. These assumptions are restrictive since, for example, the isolated Kerr black hole 
does not admit conformally flat slices~\cite{KerrNotFlat}, and consequently the Bowen-York solution 
cannot be used to construct 
black holes with spin higher than $S/M_{\rm ADM}^2 = 0.928$~\cite{MaxBYSpin,LovelaceID}.
Other examples include the use of quasiequilibrium assumptions for
constructing ID for binary systems (such as approximate
helical Killing vectors or the like, and approximate hydrostatic equilibrium
for any matter in the system); see, for 
example,~\cite{1993PhRvD..47.1471C,Cook:1994va,1996PhRvD..54.1317W,1997PhRvL..79.1182B,Bonazzola:1997gc,QE-XCTS-1,PhysRevD.65.044020,PhysRevD.65.044021,Tichy:2003qi,PhysRevD.70.104016,Ansorg:2004ds,Ansorg:2005bp,2004PhRvD..70d4044S,lorene_bhns_4,sxs_bhns,PhysRevD.74.041502}.
This serves as a good approximation for astrophysically 
motivated quasicircular inspiral sufficiently far from merger, though it is not valid for 
eccentric mergers (except possibly near the turning points of the orbit~\cite{PhysRevD.77.044011}) 
or the ultrarelativistic scattering problem. Many of these studies made further
simplifying assumptions, such as conformal flatness, which does not have
an astrophysical motivation. Attempts to supply more realistic conformal initial
data include superposition of isolated black hole spacetimes~\cite{PhysRevD.66.024047,Hannam:2006zt,PhysRevD.59.024015,PhysRevLett.85.5500,
lovelace_cqg,LovelaceID,Liu:2009al}, and in addition using post-Newtonian solutions~\cite{PhysRevD.67.064008,PhysRevD.73.124002} and matched
asymptotic expansions to supply an initial outgoing radiation 
field~\cite{Yunes:2005nn,JohnsonMcDaniel:2009dq,tichy}. Using superposed data allowed evolution
of binary black holes in quasicircular orbits with spins exceeding the Bowen-York limit~\cite{LovelaceEvo}.
A further alternative approach, initially applied to binaries including neutron stars, involves
solving the full Einstein-Euler system of equations with a waveless and/or near-zone helical
symmetry approximation~\cite{2004PhRvD..70d4044S,2004PhRvD..70l9901S,Uryu:2005vv,PhysRevD.78.104016,2009PhRvD..80l4004U}.

Since our goal is to have a more general purpose numerical initial data solver that can
be used for a range of applications, as outlined in the first paragraph, we
will use the CTS formalism with arbitrary conformal metric and other free data
to be chosen as needed for the particular application. For our first version
of the code, as presented here, we restrict to four-dimensional, asymptotically flat spacetimes,
with application to BCO interactions. For the free data we use superposed, boosted single
CO spacetimes. At large separation this is a good approximation for the physical
metric of dynamical capture binaries and the ultrarelativistic scattering problem,
and the nonlinear corrections from solving the CTS equations are small. For quasicircular
binaries, again at large separation this is a good approximation. However, unlike the
scattering problems, at practical (because of limited computational resources) initial
separations to allow evolution through merger, the simple superposition we
use at present will not give improved astrophysically relevant ID compared to current
quasiequilibrium approaches. Compared to existing studies using superposed
data, a couple of novel aspects about our work is 
we include the matter and metric in the superposition of COs involving fluid stars 
(as opposed to solving the Euler equations on a flat background in the studies mentioned earlier,
or conformal to a single black hole solution~\cite{sxs_bhns})
and the consideration of ultrarelativistic initial boosts with Lorentz factors up to $10$. 

Another notable aspect of this work is how we handle black hole singularities.
Most existing approaches either use some form of boundary condition
on a trapped surface on or inside each black hole (see, for example,~\cite{1989fnr..book...89Y,1993PhRvD..47.1471C,PhysRevD.70.104016}),
or use a slice that maps the interior region of the computational domain for each black hole
to either part (so-called ``trumpets''~\cite{PhysRevLett.99.241102,PhysRevD.78.064020,PhysRevD.80.124007,Immerman:2009ns}) 
or all (``punctures''~\cite{Brandt:1997tf}) of a different asymptotically flat region spanned by 
an Einstein-Rosen bridge (for a novel variant that does not
require separation of the metric into a background piece and conformal factor see~\cite{Baumgarte:2012dt}).
Here we follow an alternative approach where some distance inside the apparent horizon of
each black hole we replace the vacuum interior with an (unphysical) distribution of
stress-energy to regularize the interior metric. This is similar to a 
``stuffed black hole"~\cite{PhysRevD.57.2397} or the 
``turduckening'' evolution scheme~\cite{turducken1,turducken2} (see also~\cite{PhysRevD.54.R5931}).  
However, since we use excision
to subsequently evolve the initial data, with the excision surface chosen to 
entirely contain the unphysical matter, here it is merely a device to set up a
simple initial data problem without explicit interior boundary conditions or 
singularities. Note, however, that if we were to solve the ID on a domain with
traditional excision surfaces inside each black hole, we would (assuming a well-posed
elliptic problem) obtain the same solution exterior with appropriate excision boundary
conditions, though the mapping between some unphysical interior and appropriate
boundary conditions would be nontrivial and in general nonunique.

An outline of the rest of the paper is as follows.
In Sec.~\ref{secComputational} we review the CTS formulation, 
describe our method for choosing the metric and fluid free data, outline the scheme for regularizing black hole solutions,
and describe how we numerically solve the constraint equations using a multigrid solver. 
In Sec.~\ref{secApplications} we present examples of initial data obtained 
with our solver for quasicircular, eccentric, and ultrarelativistic mergers of compact objects. 
Finally, we comment on our results and discuss possible future improvements in Sec.~\ref{secConclusions}.
In the Appendix we give some details on how we treat mesh refinement boundaries in our multigrid algorithm. 
We use geometric units where Newton's constant $G=1$ and the speed of light $c=1$.

\section{Computational methodology}
\label{secComputational}

\subsection{Conformal thin-sandwich equations}
\label{subsecCTSeqn}
To formulate the initial data problem for general relativity, we start by foliating spacetime with a family of spacelike 
hypersurfaces $\Sigma_t$ parametrized by $t$. The normal vector to these surfaces $n^{\mu}$ and the generator of time
translations $t^\mu$ satisfy
\beq  t^{\mu} = \alpha n^{\mu} + \beta^{\mu}, \eeq
where $\alpha$ is the lapse and $\beta^{\mu}$ is the shift, which is tangent to $\Sigma_t$ ($n_{\mu} \beta^{\mu}=0$).
We use the standard convention where Greek indices run through $\{0,1,2,3\}$ and represent the full spacetime
coordinates, while Latin indices run through $\{1,2,3\}$ and represent coordinates intrinsic to
a given spatial hypersurface. Using the orthogonal projection operator $\perp^{\mu}{}_{\nu} \equiv \delta^{\mu}{}_{\nu}+n^{\mu}n_{\nu}$, 
we obtain the induced metric on $\Sigma_t$, $\gamma_{ij}\equiv g_{\mu\nu} \perp^{\mu}{}_{i} \perp^{\nu}{}_{j}$,
where $g_{\mu\nu}$ is the four-dimensional spacetime metric. The line element can be written in terms of these quantities as
\beq  ds^2 = -\alpha^2 dt^2+\gamma_{ij} (dx^i + \beta^i dt) (dx^j + \beta^j dt) . \eeq
The extrinsic curvature of a slice $\Sigma_t$ can be written in terms of a Lie derivative as
\beq K_{ij} \equiv -\frac{1}{2} \Liec{n} \gamma_{ij} \ . \eeq 
Projecting the Einstein equations onto the hypersurface $\Sigma_t$, one obtains the constraint
equations
\begin{align}
R + K^2 +K_{ij} K^{ij} &= 16\pi E ,\label{eqHamConstraint}\\
D_j K^{ij} - D^i K &=8\pi p^i \label{eqMomConstraint},
\end{align}
where $K = \gamma^{ij} K_{ij}$, $R$, and $D_i$ are the Ricci scalar and covariant derivative associated 
with $\gamma_{ij}$, respectively, and $E$ and $p^i$ are 
the energy and momentum density as measured by an Eulerian observer, respectively.

In the language of the $3+1$ decomposition, initial data for the Einstein field equations (and any
matter evolution equations) are
a set of $20$ functions representing the components of $\alpha$, $\beta^i$, $\gamma_{ij}$, $K_{ij}$, $E$, and $p^i$ 
on the initial slice $\Sigma_t$ that together satisfy the constraints (\ref{eqHamConstraint})--(\ref{eqMomConstraint}). 
Though, in principle, there are numerous conceivable ways of coming up with consistent initial data, it is
challenging to separate freely specifiable versus constrained degrees of freedom
in a manner where the underlying physical interpretation of the free data is transparent, and where the choice of the free data leads to 
a well-posed set of constraint equations.
The CTS method~\cite{YorkCTS} is a prescription for this separation of degrees of freedom that begins with 
a conformal decomposition of the spatial metric and the extrinsic curvature.
Introducing the conformal factor $\Psi$, we define
\begin{align}
 \t{\gm}_{ij} &\equiv \Psi^{-4} \gm_{ij}, \\
 \hA^{ij} &\equiv \Psi^{10} \left(K^{ij} -\frac{1}{3} K \gm^{ij} \right) \nonumber \\ 
&= \frac{1}{2\t{\alpha}} \left[ \dot{\t{\gm}}^{ij} + \tD^i \beta^j +\tD^j \beta^i -\frac{2}{3} \t{\gm}^{ij} \tD_k \beta^k  \right]\label{eqCTS2_a} ,
\end{align}
where $\dot{\t{\gm}}^{ij}\equiv \Psi^4(\dot{\gm}^{ij} - \frac{1}{3}{\gm}^{ij}\gm_{kl}\dot{\gm}^{kl})$
is defined to be traceless, the overdot indicates a time derivative,
$\t{\alpha} \equiv \Psi^{-6} \alpha$, and $\t R$ and $\t D_i$ are the Ricci scalar and covariant derivative associated 
with $\t{\gm}_{ij}$, respectively. With these definitions we
can rewrite~(\ref{eqHamConstraint}) and~(\ref{eqMomConstraint}) in the CTS form as
\begin{align}
\tilde{D}_i \tilde{D}^i \Psi -\frac{{\tilde R}}{8}  \Psi
	+ \frac{1}{8} \hat{A}_{ij} \hat{A}^{ij}\Psi^{-7}
	 - \frac{K^2}{12}  \Psi^5 &= - 2\pi \Psi^{-3} {\tilde E} \label{eqCTS1}\\
	 \tilde{D}_j \hat{A}^{ij}
  - \frac{2}{3} \Psi^6 \tD^i K &= 8\pi {\tilde p}^i \label{eqCTS2}  
\end{align}
with $\t{p}^i\equiv\Psi^{10} p^i$, $\t{E} \equiv \Psi^8 E$. Initial data is obtained by solving this system of four elliptic equations 
for $\Psi$ and $\beta^i$ [upon substitution of (\ref{eqCTS2_a}) into (\ref{eqCTS2})], where
$\tilde{\gamma}_{ij}$, $\dot{\tilde{\gamma}}^{ij}$, $K$, $\t{\alpha}$, $\t{E}$, and $\t{p}^i$ are the ``free data'' that can 
be freely specified to reflect the physical system under investigation. 

\subsection{Superposed free data}
\label{sp_free_data}

Under the conformal thin-sandwich method one is free to choose any
values for $\tilde{\gamma}_{ij}$, $\dot{\tilde{\gamma}}^{ij}$, $\tilde{\alpha}$, $K$, $\t{E}$, and $\t{p}^i$
for which a solution can be found.  In this section we outline our method for 
determining this free data in order to construct initial data representing binary systems.  The basic
idea is as follows. Since solutions to the Einstein equations representing isolated compact objects 
(black holes, stars, etc.) are well known, and since if the separation between the objects is not too small
the solution describing two compact objects is well-approximated by superposing the two isolated solutions, we
therefore set our free data using such a superposed solution and then solve the constraint
equations in order to obtain the nonlinear correction.

There are many ways to combine the metrics representing isolated compact objects.  The 
method we use is based on the $3+1$ splitting.  Let $\gamma_{ij}^{(1)}$, $\dot{\gamma}_{ij}^{(1)}$,
$\alpha^{(1)}$, and $\beta^{i(1)}$ represent the spatial metric, time derivative of the spatial metric,
lapse, and shift, respectively, of the first isolated solution (e.g. a boosted black hole or neutron star solution) 
and similarly for the second isolated solution.  Then, we construct the following quantities:

\begin{eqnarray}
\gamma_{ij}^{(\rm sup)} &=& \gamma_{ij}^{(1)} + \gamma_{ij}^{(2)} - f_{ij} \\
\dot{\gamma}_{ij}^{(\rm sup)} &=& \dot{\gamma}_{ij}^{(1)} + \dot{\gamma}_{ij}^{(2)} \\
\alpha^{(\rm sup)} &=& \alpha^{(1)} + \alpha^{(2)} - 1 \\
\beta^{i {(\rm sup)}} &=& \beta^{i(1)} + \beta^{i(2)}
\end{eqnarray}
where $f_{ij}$ is the flat-space metric. This particular construction will break down if $\alpha^{(\rm sup)}\leq 0$
or $\det[\gamma_{ij}^{(\rm sup)}]\leq 0$ anywhere in the domain, which would then require some other way
of combining the metrics, for example, using distance-weighted attenuation functions as in~\cite{Marronetti:2000rw}.
(In~\cite{lovelace_cqg} it was also found necessary to enforce a desired 
asymptotic falloff of the superposed metric, owing to the use of a corotating frame.)
However, these conditions are not violated for the cases considered here.
From the above quantities, we then calculate the free data we will use when solving the CTS equations from the usual relations:
\begin{eqnarray}
\tilde{\gamma}_{ij} &=& \gamma_{ij}^{(\rm sup)} \\ 
\dot{\tilde{\gamma}}^{ij} &=& -\tilde{\gamma}^{ik} \tilde{\gamma}^{jl} \left(\dot{\gamma}_{kl}^{(\rm sup)}
                              -\frac{1}{3}\tilde{\gamma}^{mn}\dot{\gamma}_{mn}^{(\rm sup)}\tilde{\gamma}_{kl} \right)  \\ 
                              \tilde{\alpha} &=& \alpha^{(\rm sup)}\\
K &=&   \frac{1}{2\tilde{\alpha}}(2\partial_i \beta^{i {(\rm sup)}}+\dot{\tilde{\gamma}}^{ij}\tilde{\gamma}_{ij} \nonumber \\
  &\ & \ \ \ \ + \tilde{\gamma}^{ij}\beta^{k {(\rm sup)}}\partial_k \tilde{\gamma}_{ij}) .
\end{eqnarray}

For initial data with matter we use a similar method.  We set $\tilde{E}$ and $\tilde{p}^i$ by
superposing the energy and momentum density of the two objects (we do not consider situations where
they would both be nonzero at the same point).
For some cases (in particular for the ultrarelativistic boosts), we rescale the momentum density so that 
its magnitude with respect to the superposed metric $\tilde{\gamma}_{ij}$ is equal to the magnitude of the original 
momentum density with respect to the metric of the isolated object ($\gamma_{ij}^{(1)}$ or $\gamma_{ij}^{(2)}$).
This ensures that $\tilde{E}^2$ and $\tilde{p}^i \tilde{p}_i$ have the same ratio as the isolated objects.
This is important since the choice of conformal scaling of the energy $\tilde{E}=E\Psi^{8}$ was designed 
to ensure that if the conformal quantities satisfy the dominant energy condition, 
$\sqrt{\tilde{\gamma}_{ij}\tilde{p}^i\tilde{p}^j}\leq \tilde{E}$, then so will the rescaled quantities
following the solution of the constraints.

\subsection{Regularizing black hole solutions}
\label{subsecRegularization}
In cases where black holes are a part of the physical system, the divergences at the black hole's singularity must be addressed. 
As discussed in the Introduction, there are several ways to deal with this issue in the initial data problem;
the approach we take here is to explicitly modify the metric of an isolated (prior to superposition) black hole solution inside 
the horizon to take a prescribed, regular form. The regularized region will not in general satisfy the vacuum constraint
equations, and to avoid a singular conformal factor and shift vector components when solving the constraints with
such background data, we introduce unphysical energy-momentum in the union of black hole interiors so that
these regions automatically satisfy the constraints, albeit with the unphysical interior matter source.

We start with a single, unboosted spinning black hole spacetime in horizon penetrating coordinates
(for the results described here we use the harmonic form of Kerr derived in~\cite{CookScheel},
though we have also tried it using Kerr-Schild coordinates without difficulty), so that the only
divergences in the metric components are well within the horizon.
We then choose a surface that encloses the singular region, yet is strictly inside the event horizon. The interior of
this surface we call the {\em regularization region}. Outside the regularization region we do not modify the 
metric. Inside, there are many conceivable ways to alter the metric to eliminate the divergences. The simple
approach we take is to promote the black hole mass $M$ and spin $a$ constants to functions of space and smoothly
decrease them from their bare values at the regularization surface to zero at some surface 
interior to this.

Specifically, we introduce a regularization function
\begin{equation}  
f_{\rm reg}(x) = \left\{
     \begin{array}{cc}
       1, &  x>1, \\
       x^3\left(6x^2-15x+10\right), &   1>x>0, \\
       0, &  0>x,
     \end{array}
   \right.
\end{equation}
chosen to be twice continuously differentiable so that the consequent unphysical energy is well behaved. 
We use a Cartesian grid\footnote{Note that in the harmonic coordinates of~\cite{CookScheel} 
only the region with $r_{K}>M$ is represented on the Cartesian grid
$r\ge0$, where $r_{K}$ is the radial coordinate of the the metric in ingoing null Kerr form. Hence the physical 
singularity is not on the grid; however, the metric components are discontinuous at $x=y=z=0$ ($r_K=M$), and hence regularization 
is still required.}
and define $r^{(E)}(x,y,z)=\sqrt{x^2+y^2+z^2}$ as the Euclidean radius for a point with coordinates $x,y,z$. 
We then replace the mass $M$ and the spin parameter $a$ with $\xi(x,y,z) M$ 
and $\xi(x,y,z) a$, respectively, in all the metric components, where 
\begin{equation}
 \xi(x,y,z) = f_{\rm reg}\left(\frac{(r^{(E)}(x,y,z)/r^{(E)}_+(x,y,z))-q_{\rm in}}{q_{\rm out}-q_{\rm in}}\right),
\end{equation}
$r^{(E)}_+(x,y,z)$ is the Euclidean radius for the point on the event horizon at the same angular direction 
as $(x,y,z)$, $q_{\rm out}$ defines the outer surface of the regularization region, and 
$q_{\rm in}$ is the inner surface inside of which the metric is Minkowski, with $1>q_{\rm out}>q_{\rm in}>0$.
The shape of the regularization region, namely a shrunken form of the interior
of the event horizon, was motivated by the similar volume excised during evolution
(though that is based on the apparent horizon, and the excision surface is a best-fit ellipsoid rather
than the exact shape of the apparent horizon). The particular values of $q_{\rm out}$ and
$q_{\rm in}$ are not too important (i.e., give essentially the same solutions), the only practical 
requirements being that $q_{\rm out}$ represents a surface within the excision surface we will use during
evolution and that $q_{\rm in}$ is not to be too close to $q_{\rm out}$; otherwise excessive resolution
is needed to resolve the transition.

Once we have an everywhere-regular metric, we superpose it with any other COs to construct the 
free data as described in Sec.~\ref{sp_free_data}. We then compute the unphysical energy
and momentum we will add to the regularization regions simply by evaluating~(\ref{eqHamConstraint}) and~(\ref{eqMomConstraint}) 
with the background, superposed data
\begin{align} \label{fake_energy}
E_{\rm unphys} &\equiv \frac{1}{16\pi}(R + K^2 +K_{ij} K^{ij})^{\rm (sup)}, \\
p^i_{\rm unphys} &\equiv \frac{1}{8\pi}(D_j K^{ij} - D^i K)^{\rm (sup)} .
\end{align}
$E_{\rm unphys}$ and $p^i_{\rm unphys}$ are then added to $\tilde{E}$ and $\tilde{p}^i$
within the regularization regions, 
and we can then solve the CTS equations as usual without any additional special treatment of 
these regions. It is also possible to calculate the unphysical 
energy-momentum before the superposition and add 
$E_{\rm unphys}$ and $p^i_{\rm unphys}$ directly. The former method gives a small discontinuity
of $E_{\rm unphys}$ and $p^i_{\rm unphys}$ at the boundary of the regularization region,
whereas the latter one gives continuous quantities. 
Either approach leads to similar results, but the former gives more rapid relaxation of
the elliptic equations and is the choice for the cases presented here.

During evolution, we choose black hole excision surfaces that
entirely contain the regularization regions and unphysical matter.
Thus, one can think of the unphysical matter as serving as a proxy for what
would otherwise be boundary conditions for $\Psi$ and $\beta^i$ 
on excision surfaces. Given a solution to the constraints with regularized
interiors it is trivial to read off what the equivalent (Dirichlet) boundary
conditions would have been, though the inverse problem of mapping some
set of desired boundary conditions to interior sources is less trivial and
likely not well posed in general.

\subsection{Fluid solutions}
\label{subsecFluid}
For the applications with (physical) matter considered here we use Tolman-Oppenheimer-Volkov (TOV) 
star solutions in isotropic coordinates to construct
the metric free data quantities as well as $\tilde{E}$ and $\tilde{p}^i$.  Such solutions are derived by
assuming a relationship between the pressure and density $P(\rho)$, e.g. as given by a polytropic condition.
Once the constraint equations have been solved and $E$ and $p^i$ found, we determine 
the new density and pressure profiles
using this same relationship and solving the equation
\begin{equation}
(E+P(\rho))(E-\rho)-p_i p^i=0 
\end{equation}
for $\rho$, which follows directly from the expressions for the energy and momentum density of a perfect fluid.  
For the applications considered here, we do not explicitly impose any additional constraints on the
fluid quantities (e.g. that the fluid be in hydrostatic equilibrium).  We leave that to future extensions.  

\subsection{Multigrid elliptic solver}
\label{subsecMG}
To numerically solve the CTS equations we discretize~(\ref{eqCTS1}) and~(\ref{eqCTS2})  using 
standard second-order finite difference operators and solve 
them using a full approximation storage implementation of the multigrid algorithm with adaptive mesh refinement (AMR) 
as described in~\cite{Pretorius:2005ua}.  
A multigrid algorithm is characterized by a smoothing operation and by a choice of restriction and 
prolongation operators. We use Newton-Gauss-Seidel relaxation for smoothing, and half-weight restriction
and linear interpolation for the restriction and prolongation operators, respectively. These latter operators require special 
treatment on mesh refinement boundaries which we outline in the Appendix.  
Unlike in the evolution code, we do not use a compactified coordinate system extending to spatial
infinity.  Rather, the initial data numerical grid extends to a large but finite radius. This is to
avoid numerical problems attributable to large Jacobian factors needed in compactification which become 
especially problematic near the corners of the boundary.  
At the outer boundaries we impose boundary conditions that $\Psi=1$ and $\beta^i=\beta^{i {(\rm sup)}}$.
Since the use of mesh refinement enables us to put the outer boundary far away from the compact objects, these
boundary conditions can be made sufficiently accurate compared to numerical error 
(though for future applications they could also be replaced with, e.g., Robin boundary conditions).
Any points outside this domain on the evolution grid are initialized via extrapolation, assuming
a leading order $1/r$ approach to an asymptotically flat spacetime.

For some applications we wish to solve for initial data with axisymmetry.  To efficiently solve the constraint
equations in these situations we have implemented a modified version of the Cartoon method~\cite{Cartoon} similar to that used 
in~\cite{Pretorius:2004jg}.  Letting the $y$ axis be the axis of symmetry, we restrict our computational domain to a subset of 
the half-plane $(x,y)\in (-\infty,\infty)\times [0,\infty)$.
We use the existence of an axisymmetric Killing vector to express derivatives in the $z$ direction in terms of 
derivatives in the $x$ and $y$ directions.  On the $y$ axis we impose regularity, which gives the following conditions
for the constrained variables: $\partial_y \Psi = 0$ and $\partial_y \beta^x=\partial_y \beta^y=\beta^z=0$.

\section{Applications}
\label{secApplications}

\subsection{Quasicircular binary black holes}
\label{subsecQuasi}
As a first application of our technique, we generate and evolve ID for the (approximate) quasicircular inspiral of two nonspinning,
equal-mass black holes. Our present method for providing free data is not designed 
to easily give initial data for quasicircular inspiral (though presumably with sufficient
fine-tuning of the boost vectors this could be achieved), and this basic example is mainly to provide
a relatively low eccentricity binary, a couple of orbits before merger, for comparison to past studies. Specifically,
we are interested in seeing how close the masses, etc., of the black holes obtained following the 
solution of the constraints are to the corresponding parameters used in constructing the free data,
and how much ``spurious" gravitational radiation is present in the initial data.

\begin{figure}
\begin{center}
\includegraphics[width=3.2in,clip=true,draft=false]{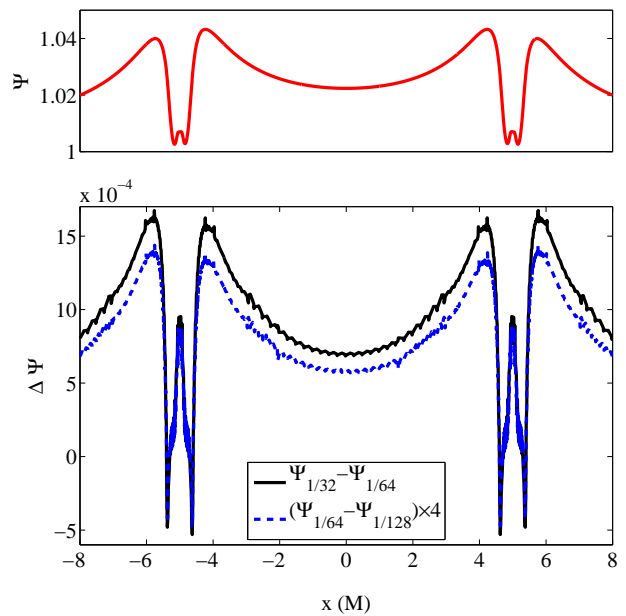}
\end{center}
\caption{
The conformal factor $\Psi$ from BH-BH ID.
Upper: $\Psi$ on the $x$ axis which lies on the orbital plane and goes through the centers of the black holes.
Lower: Differences in $\Psi$ with resolution on the $x$ axis, scaled assuming second-order convergence.
\label{ID_psi_convergence}
}
\end{figure}

\begin{figure}
\begin{center}
\includegraphics[width=3.2in,clip=true,draft=false]{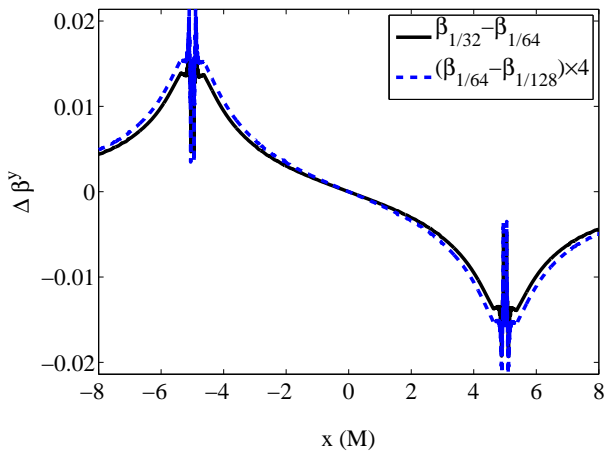}
\end{center}
\caption{
Differences in the shift component $\beta^y$ with resolution on the $x$ axis from BH-BH ID, 
scaled assuming second-order convergence.
\label{ID_beta_convergence}
}
\end{figure}

\begin{figure}
\begin{center}
\includegraphics[width=3.2in,clip=true,draft=false]{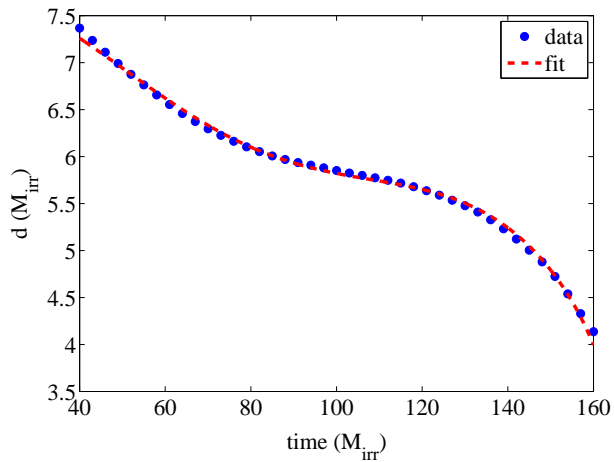}
\end{center}
\caption{
Coordinate separation of the centers of the two black holes fitted to a function
$(A-B(t-t_0))^{1/4}+C \cos(\omega (t-t_0) + \phi)$. This function combines the decaying orbit attributable to quadrupole radiation
with the effects of eccentricity, given by $e=C/d(t=t_0) \approx 0.05$. Because of early-time gauge
effects (a transition from harmonic to damped harmonic gauge) we exclude the
first $t=40M_{\rm irr}$ from the fit.
\label{bbh_eccentricity}
}
\end{figure}

\begin{figure}
\begin{center}
\includegraphics[width=3.2in,clip=true,draft=false]{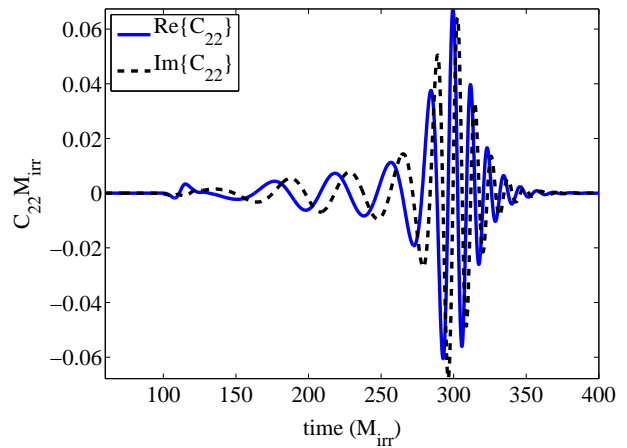}
\end{center}
\caption{
The real and imaginary components of the $l = 2$, $m = 2$ spin-weight $-2$ spherical harmonic of 
$r\Psi_4$ extracted at a radius of $105M$. Time is measured from the beginning of the simulation.
\label{bbh_gwave}
}
\end{figure}

For the initial data, we use free data set by superposing two boosted nonspinning
equal mass black holes at a coordinate separation of $10M$, where $M$
is the sum of the isolated black hole masses
(which in general will be different from the irreducible masses of the black holes once the constraint equations are solved).
The black holes are given purely tangential boost velocities chosen so that, when 
evolved, the black holes undergo a few orbits with monotonically decreasing proper separation.
The initial data grid extends to $\pm2048M$ in all three directions. For convergence studies of the initial data solver,
we use three base grid sizes of $33^3$, $65^3$, and $129^3$ and $12$ levels
of mesh refinement with identical grid structures in each case.
As expected, the conformal factor and shift vector exhibit second-order convergence as shown in Figs.~\ref{ID_psi_convergence} and 
\ref{ID_beta_convergence}, as does the residual of~(\ref{eqHamConstraint}) and~(\ref{eqMomConstraint}).
For evolution, we use the highest resolution initial data.
The ID is evolved using the generalized harmonic formulation of the field equations, choosing
harmonic coordinates at $t=0$ and transitioning to a damped harmonic gauge as described in~\cite{code_paper}.
The eccentricity is estimated to be $e\approx 0.05$ based on the evolution of the coordinate distance 
between the centers of the apparent horizons as shown in Fig.~\ref{bbh_eccentricity}. 
Though the orbital eccentricity could presumably be reduced further by tuning the initial velocities using methods such as 
the one proposed in~\cite{reducing_ecc} or using the post-Newtonian approximation as in~\cite{PhysRevD.77.044037}, 
we did not attempt to do so for this basic comparison.
Because of corrections from solving the constraints, the sum of the masses of the isolated black holes 
whose spacetimes we superpose $M$ is different from the sum of irreducible masses computed from their 
apparent horizons at the beginning of the evolution $M_{\rm irr}$. 
For this particular case $M_{\rm irr}/M=1.21$.
The ID is constructed using free data with nonspinning black holes,
and the initial spin calculated from the apparent horizons is zero to within truncation 
error ($|S/M_{\rm BH}^2| <6\times10^{-3}$). The ratio of the irreducible mass of the final black 
hole after the merger to the sum of the irreducible masses of the initial black holes is 
$M_{{\rm irr},f}/M_{\rm irr}=0.885$, and the dimensionless spin parameter of the final black hole 
is $a_f/M_f=0.678$. Both of these values are in good agreement (considering the mild initial
eccentricity here) with the high accuracy results of
$0.88433$ and $0.68646$, respectively, from~\cite{high_acc_bbh}.   
In Fig.~\ref{bbh_gwave} we show the gravitational waves from the BH-BH merger.   
The initial spurious part of the signal is of comparable magnitude to other ID 
approaches that do not attempt to include gravitational waves from the prior
inspiral; see, for example,~\cite{tichy}.

\subsection{Eccentric compact object mergers}
\label{subsecEcc}
As another application of this technique, we consider constructing initial data describing a dynamical capture
BH-NS binary.  We set the free data using a boosted harmonic black hole solution and a neutron star with the HB 
equation of state~\cite{jocelyn}. Let $M$ be the sum of the masses of the isolated black hole and neutron star. 
We construct initial data for a 4:1 BH-NS binary by setting the boost velocities to correspond to a Newtonian orbit 
with eccentricity $e=1$ and periapse distance $r_p= 5 M$ at various initial separations $d$.
We keep the mass and spin that we use for the black hole component of the free data fixed at $0.8M$ and $-0.4M$, 
respectively (where the negative sign indicates that the spin is retrograde with respect to the orbital angular 
momentum) and the mass of the neutron star component of the free data fixed at $0.2M$.  The spin and masses will 
receive corrections from solving the constraint equations and with decreasing $d$ these will differ more and more 
from the input parameters of the free data.  The input parameters can, of course, be tuned 
to achieve desired values in the final solution. However, since here we are mainly interested in quantifying this 
difference, we keep them fixed.  We use a grid extending from $-1600 M$ to $1600 M$ in each dimension where the 
base level is covered by $257^3$ points and there are nine additional levels of mesh refinement, each with a 
refinement ratio of two. We solve for data with initial separations $d/M=$ 15, 25, and 50. In Table~\ref{bhns_table} 
we show the maximum difference of the conformal factor from unity as well as the actual ADM (Arnowitt-Deser-Misner) 
mass, black hole mass and spin, neutron star rest mass, and induced neutron star density oscillations for these three different 
separations.  We can see that even at a separation of 15 $M$ the difference between input and final parameters 
is small---at the level of a few percent. At such separations, however, the oscillations induced in the neutron 
star by the initial setup become large.  This problem could be remedied by adding additional constraints to the matter, 
for example, requiring it to satisfy an equilibrium version of the Euler equations. 

\begin{figure}
\begin{center}
\includegraphics[width=3.2in,clip=true,draft=false]{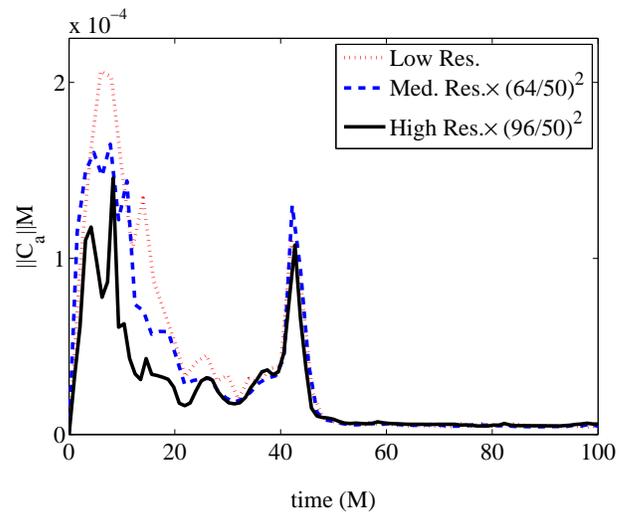}
\end{center}
\caption{
The $L^2$ norm of the constraint violation, $C_a \equiv H_a-\Box x_a$, in units of $1/M$ for the $d=15M$ BH-NS merger in the
$100M \times 100M$ region around the center of mass in the equatorial plane (i.e. $\sqrt{\int\|C_a\|^2 d^2x / \int d^2x}$ ).
This is shown for low, medium, and high resolutions where the latter two are scaled assuming second-order convergence.
\label{ecc_cnst}
}
\end{figure}

\begin{figure}
\begin{center}
\includegraphics[width=3.2in,clip=true,draft=false]{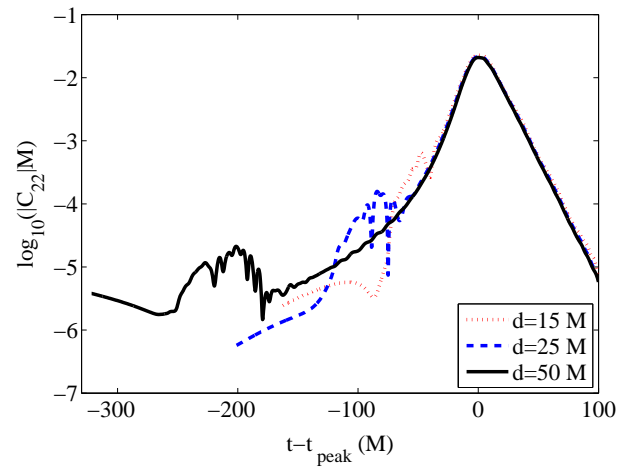}
\end{center}
\caption{
The log of the magnitude of the $l=2$, $m=2$ spin-weight $-2$ spherical harmonic of $r\Psi_4$ for BH-NS simulations with different
initial separations $d$.  The value of $\Psi_4$ was extracted on a sphere of radius $100 M$ and is shown starting
at the beginning of the simulation and continuing past merger.  The waveforms have been aligned so that the 
peaks occurs at time 0.
\label{ecc_gwave}
}
\end{figure}

We evolve the initial data past merger using the same methods, gauge, and three different resolutions as 
in~\cite{bhns_astro_paper}. Unless otherwise stated, all quantities are from the high resolution runs.
In Fig.~\ref{ecc_cnst} we show the norm of the constraints throughout the evolution of the $d=15M$ ID at the 
different resolutions. The single highest, resolution ID is used for all evolution runs, so the fact that
evolution constraints are converging to zero indicates that the truncation error of the ID is at least
as small as that of the highest resolution evolution.
In Fig.~\ref{ecc_gwave} we plot the amplitude of the gravitational waves measured from the three 
different evolutions to show the amount of spurious gravitational radiation this method of constructing 
ID introduces.  The level of spurious gravitational radiation decreases with increasing separation and in all 
three cases is small---an order of magnitude or more below the physical signal of interest.  After the passage of 
the spurious gravitational radiation, the gravitational wave signal from all three initial
separations is approximately the same, though there are small differences owing to the changes in parameters indicated 
in Table~\ref{bhns_table}, and because we are starting the systems along different points of a Newtonian trajectory.  

\begin{table*}
\begin{center}
{\small
\begin{tabular}{ c c c c c c c }
\hline\hline
$d/M$    & $\rm{max}(|\Psi-1|)$ &  $M_0/M_{0,\infty}$ & $M_{\rm BH}/M$ & $a_{\rm BH}/M$ & $M_{\rm ADM}/M$ & $\rho_{\rm oscill.}$ (\%)\\
\hline
15   &  0.0155   &    1.077   &   0.832      &   $-0.398$     &  1.051   &  14.3\\
25   &  0.0092   &    1.049   &   0.818      &   $-0.402$     &  1.030   &  9.0 \\
50   &  0.0046   &    1.028   &   0.808      &   $-0.399$     &  1.017   &  4.5 \\
\hline
\end{tabular}
}
\caption{
Characteristics of BH-NS initial data with Newtonian orbital parameters $r_p=5 M$ and $e=1$ with 
three  initial coordinate separations $d$. Here $\rm{max}(|\Psi-1|)$ is the maximum
deviation over the entire domain of the conformal factor from the background free-data value of unity,
$M_0/M_{0,\infty}$ is the rest mass of the neutron star 
compared to its isolated rest mass, $M_{\rm BH}/M$ and $a_{\rm BH}/M$ are the black hole
mass and spin parameters measured from the apparent horizon relative to the initial
total mass $M$ of the free data, $M_{\rm ADM}/M$ is the relative ADM mass
of the solution, and $\rho_{\rm oscill.}$ is the relative magnitude of the oscillation 
in time of the maximum rest mass density of the neutron star induced by the ID construction.
}
\label{bhns_table}
\end{center}
\end{table*}

\subsection{Ultrarelativistic initial data}
\label{subsecUR}
As a final application, we consider
the problem of specifying ID for ultrarelativistic collisions. The study of the collision of objects where
kinetic energy dominates the dynamics of the spacetime is of considerable interest to super-Planck scale particle
collisions, as arguments suggest classical Einstein gravity will be adequate to describe the
process~\cite{1987PhLB..198...61T,1999hep.th....6038B,2002JHEP...06..057K}. The hoop conjecture~\cite{thorne_hoop}
predicts that the generic outcome of a sufficiently ultrarelativistic collision will be black hole formation,
and this, together with suggestions of a tera-electron-volt Planck scale~\cite{1998PhLB..429..263A,1999PhRvL..83.3370R}, imply that, if such a 
scenario describes nature, the Large Hadron Collider or cosmic ray collisions with Earth
could produce black holes~\cite{2001PhRvL..87p1602D,2002PhRvD..65e6010G,2002PhRvL..88b1303F}. 
Though to date no signs of black hole production have been observed~\cite{Heros:2007hy,Chatrchyan:2012taa}, 
the nature of the kinetic energy dominated
regime in general relativity is of interest in its own right and has largely been unexplored.

Initial data describing such systems will be far from equilibrium, and one cannot assume that the solution is time symmetric
or quasistatic.    
It is instructive to recall the Aichelburg-Sexl~\cite{Aichelburg:1970dh} solution describing a gravitational shock wave. 
The solution can be obtained from 
a boosted Schwarzschild solution by simultaneously taking the mass to zero and the boost parameter to infinity, while 
keeping their product constant and finite.  Two such oppositely boosted solutions can be superposed to obtain a new solution that is 
valid up until collision.  Though it is not clear how applicable this is to the nonlimiting case, this suggests
that superposition may be a good approximation to describing such spacetimes. 

Here we consider the specific example of the setup for a head-on collision of two fluid star solutions.
We use the method described in Sec.~\ref{sp_free_data} to construct free data from two $\Gamma=2$ polytropic TOV star solutions 
that have unboosted mass $M_{*}$ and a compactness (ratio of mass to radius) of $C=0.01$.  The stars are 
boosted toward each other with boost factor $\gamma=10$. 
We consider a sequence of solutions at various initial coordinate separations $d$.  
We take advantage of the axisymmetry of the problem and use $[-2000M,2000M]\times[0,2000M]$ where $M\equiv 2\gamma M_{*}$
as our computational domain.  The base level is covered by $1025\times513$ points, and there are nine additional levels of mesh
refinement.  To test convergence we also consider two lower resolutions with grid spacing $2$ and $4/3$ times
as coarse.  

\begin{figure}
\begin{center}
\includegraphics[width=3.5in,clip=true,draft=false]{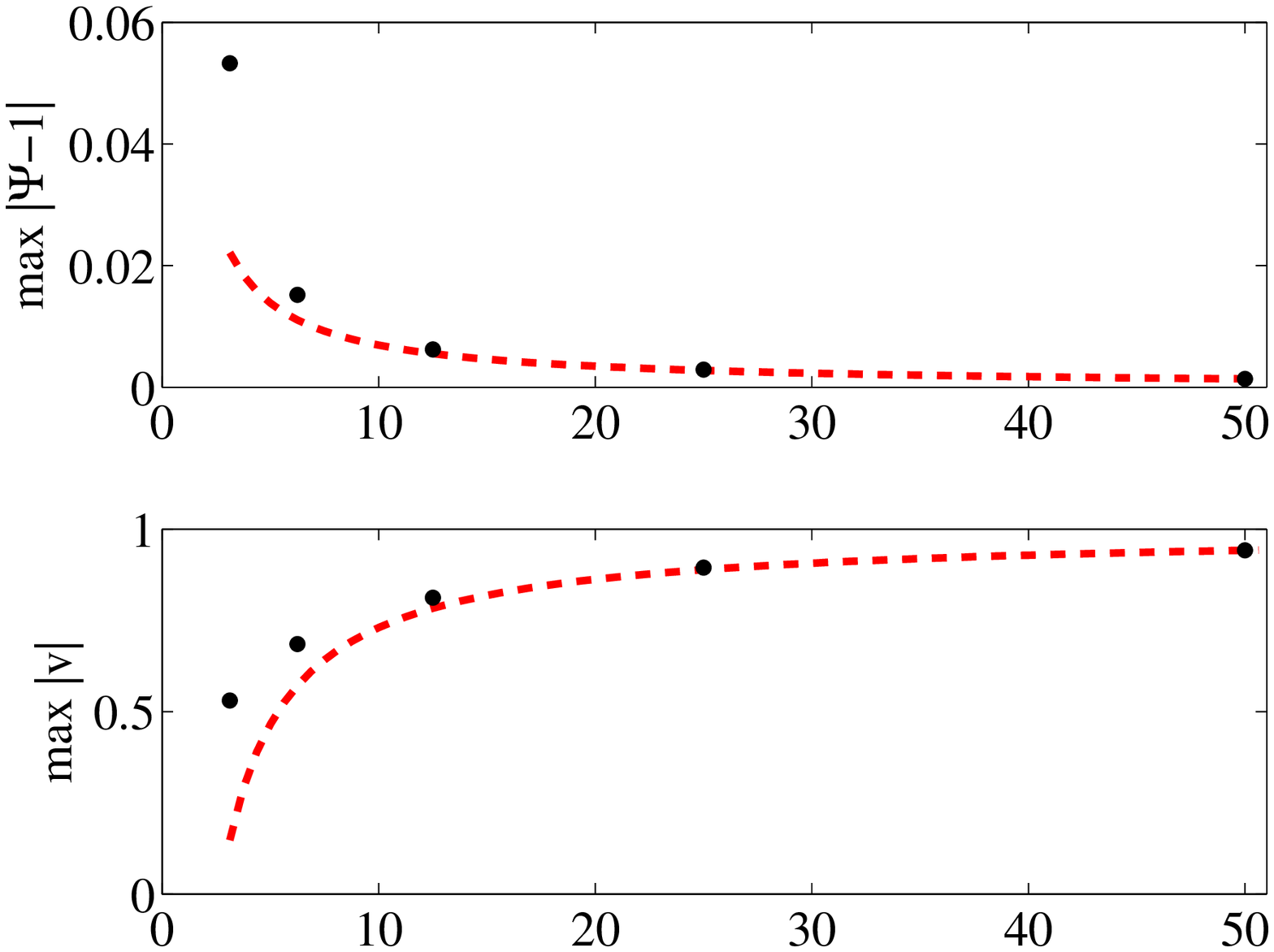}
\includegraphics[width=3.5in,clip=true,draft=false]{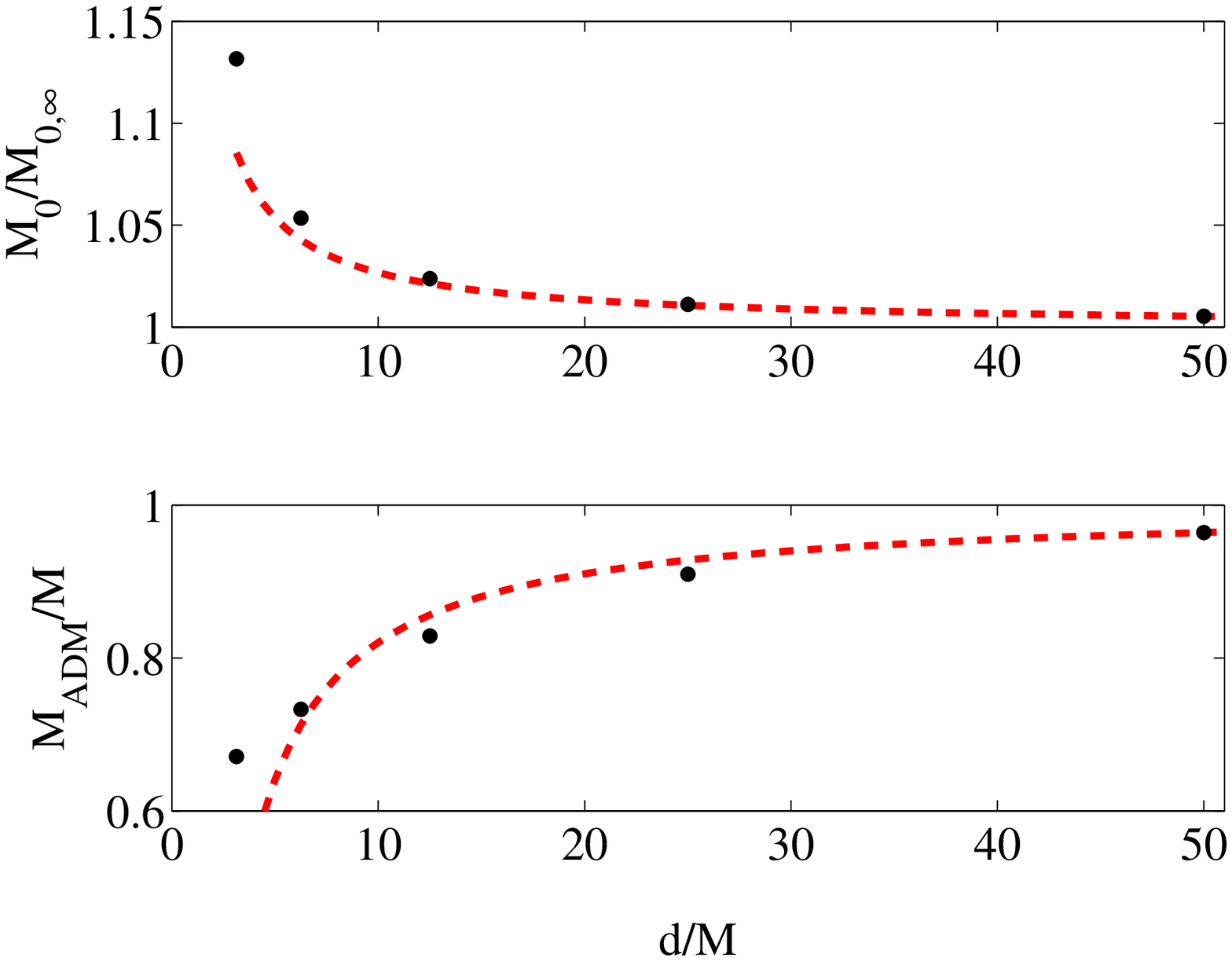}
\end{center}
\caption{
Various quantities from ultrarelativistic collision ID with $\gamma=10$ made using the 
superposition method for constructing free data.  From top to bottom the quantities 
shown are the maximum (over the entire domain) difference of the conformal factor 
from unity, the maximum coordinate velocity of the fluid, the total rest mass, 
and the ADM mass.  All quantities are shown as a function
of $d$, the coordinate separation between the two stars.  For all these cases the maximum 
of $|\Psi-1|$ occurs for values of $\Psi$ that are less than unity.
One might expect these quantities to approach their infinite separation limits as
$1/d$ for large $d$; the dotted lines show
such $1/d$ curves for each quantity matched to the $d/M=50$ point.
\label{ur_sp}
}
\end{figure}

Using the method for specifying free data 
described in Sec.~\ref{sp_free_data}, as $d\rightarrow \infty$ we expect the corrections
from solving the constraints will go to zero: $\Psi \rightarrow 1$, the magnitude of the coordinate velocities 
of the stars $|v|$ will approach $\sqrt{1-\gamma^{-2}}$, the ADM mass $M_{\rm ADM}$ will approach $M$,
and the total rest mass $M_0$ will approach the sum of the rest masses of the isolated stars $M_{0,\infty}$.
In Fig.~\ref{ur_sp} we show how all these quantities change with coordinate separation.  We can see that
it is possible to solve for ID where the stars are quite close together, though the corrections
become large, and in particular the ADM mass decreases quite significantly.

To give an indication of the numerical errors
on these quantities, we can compare the values obtained at the highest resolution to the Richardson extrapolated
values using all three resolutions. For example, for the smallest separation $d=1.56M$, 
we have $\rm{max}(|\Psi-1|) = 0.05326$ (0.05325) and $\rm{max} |v|=0.530149$ (0.530153) where the values in parentheses
are the Richardson extrapolated quantities (which are consistent with approximately second-order convergence).

\begin{figure}[b]
\begin{center}
\includegraphics[width=3.5in,clip=true, draft=false]{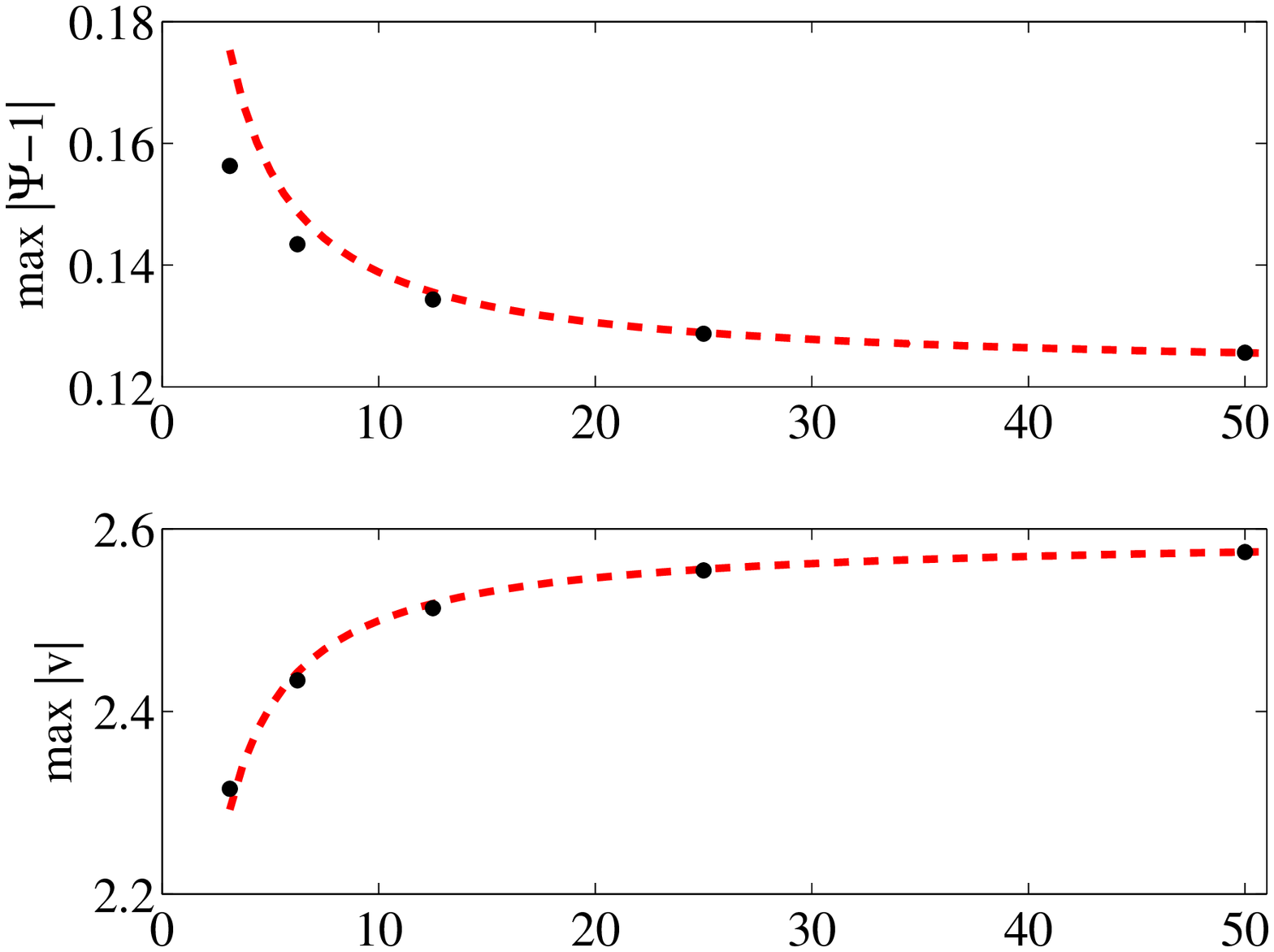}
\includegraphics[width=3.5in,clip=true, draft=false]{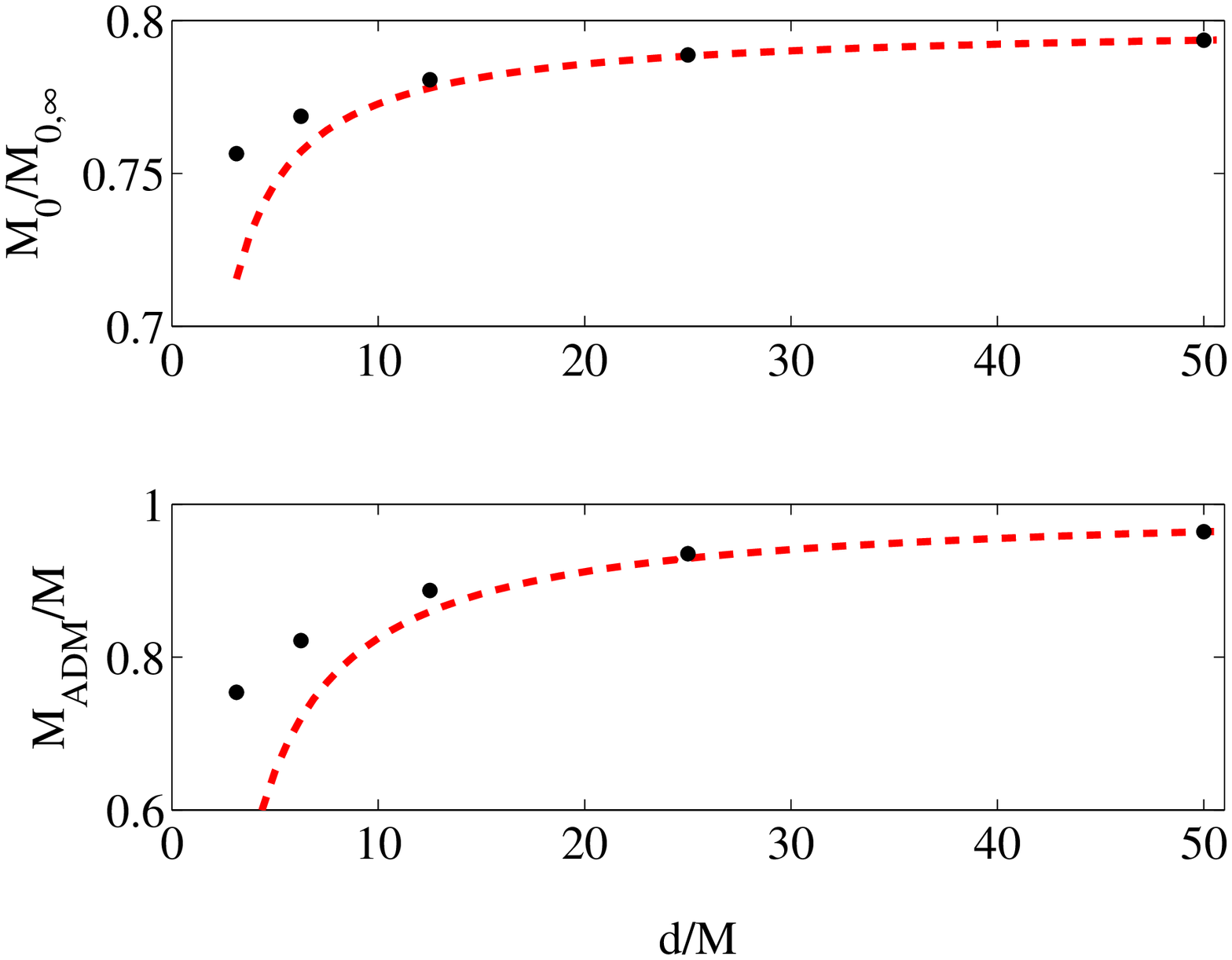}
\end{center}
\caption{
Same as Fig.~\ref{ur_sp} but with conformally flat data.
For all these cases the maximum of $|\Psi-1|$ occurs for values of $\Psi$ that
are greater than unity.
\label{ur_cf}
}
\end{figure}

We can compare the above method of constructing free data for this case to a conformally flat method.
Specifically, we set all the metric free data quantities to their flatspace values and set $\tilde{E}$ 
and $\tilde{p}^i$ for each star to a special-relativistically boosted density and pressure profile taken from 
the TOV solution.
In Fig.~\ref{ur_cf} we show the same quantities as in Fig.~\ref{ur_sp}
but using this conformally flat method.  In this case the corrections from solving the constraints
will not go to zero with infinite separation since all the nontrivial geometry is coming from the conformal factor.
Hence the energy-momentum will be substantially rescaled at any separation.    
Also in contrast to the first method, the maximum of $|\Psi-1|$ occurs for $\Psi>1$ instead of $\Psi<1$, which
means $E$ and $p^i$ will be smaller than their conformal counterparts.
With conformally flat ID it is also possible to solve for stars close together, though, as in the previous method, 
the ADM mass decreases steeply.  It should also be noted that because of the large shift vector obtained with 
the second method, the coordinate velocity is substantially greater than one, which may make it more 
challenging to numerically evolve.

A full characterization of this ultrarelativistic collision ID requires evolution, 
which we present elsewhere~\cite{ur_nsns}. 

\section{Conclusions}
\label{secConclusions}

We have outlined a general method for constructing initial data
based on superposition and the CTS formulation of the constraint equations, and 
demonstrated the method with some example solutions. Though there are numerous
existing applications of the CTS method, and superposition has been proposed
before, some of the notable aspects of the work presented here include
adding the matter and metric of neutron stars to the prescription, regularizing the interiors
of black holes with (unphysical) matter sources, and applying it to regimes not yet studied before,
namely, initial data for generic high-eccentricity binary mergers and ultrarelativistic
collisions. For astrophysically relevant binaries we find that superposition
of single, isolated compact object solutions works well in the sense that 
nonlinear correction from solving the constraints are relatively small for 
larger initial separations, implying that superposition is a good start
to attain more astrophysically realistic initial
data (for example, by adding prior gravitational wave information as in~\cite{tichy}
to the superposed background data for quasicircular or low eccentricity inspirals).
Including neutron stars, we find that the superposition effectively induces
oscillations in the stars. This again is small for large separations and hence
a good approximation to dynamical capture binaries.  However, practical application
to low eccentricity inspirals will likely require that the CTS equations
be supplemented with some form of quasiequilibrium equations for the hydrodynamics
(as in many existing ID methods, for example~\cite{Bonazzola:1997gc,2004PhRvD..70d4044S,2004PhRvD..70l9901S}).

For the ultrarelativistic boost examples we are able to obtain
solutions to the CTS equations with superposed and conformally
flat data well into the kinetic energy dominated regime ($\gamma=10$)
for sufficiently large initial separations. At smaller separations
we are still able to obtain solutions. However, for these initial data sets 
the corrections to the metric and fluid properties become large, and it is less clear 
how to separate the total energy of the spacetime into kinetic energy, rest mass energy, etc.
This will require evolution to resolve, and we leave that to future
work. Nevertheless, given that there are few results on the
uniqueness and existence of solutions to the conformal constraint equations
beyond constant mean curvature slicing~\cite{2002LRR.....5....6R} (and in
some cases, such as the extended CTS equations~\cite{XCTS} there are known examples
of nonuniqueness~\cite{nonuniqueness}), it is interesting that we are able to 
obtain solutions in this highly nonlinear regime.

\acknowledgments
We thank Geoffrey Lovelace and Sean McWilliams for useful conversations.  
We thank Hans Bantilan and Theodor Brasoveanu for working on a prototype
of the software used here.
This research was supported by NSF Grant No. PHY-0745779 and the Alfred P. Sloan Foundation (F.P.).
Simulations were run on the Woodhen and Orbital clusters
at Princeton University as well as using XSEDE resources provided by NICS under Grant No.
TG-PHY100053. 

\appendix*
\section{Multigrid AMR interpolation}
\label{mg_amr}
A multigrid algorithm requires a restriction operator to inject quantities from finer
to coarser grids as well as a prolongation operator to interpolate corrections from
coarser grids to finer grids (see e.g.~\cite{num_rec}). 
For our multigrid algorithm we use half-weight restriction as our restriction operator.
In three dimensions half-weight restriction can be written as 
\begin{equation}
f_{\rm HW} = f_{i,j,k} + \frac{1}{12}(\Delta f_{xx} + \Delta f_{yy} + \Delta f_{zz}), 
\end{equation}
where 
\begin{equation}
\Delta_{xx} f = f_{i+1,j,k}-2 f_{i,j,k}+f_{i-1,j,k},
\end{equation}
and similarly for the $y$ and $z$ directions.  
Note that $\Delta_{xx} f$ divided by $h^2$ (where $h$ is the grid spacing) is a second-order 
approximation for $\partial^2_xf$. 
On AMR boundaries where the full stencil is not available we must modify the above expression.
For example, on a negative $x$ boundary we replace $\Delta_{xx} f$ by a right-handed second derivative stencil 
\begin{equation}
\Delta_{xx} f = 2f_{i,j,k}-5 f_{i-1,j,k}+4f_{i-2,j,k}-f_{i-3,j,k}
\end{equation}
and so on for the other directions.  This ensures not only that $f_{\rm HW}$ is a second-order representation
of $f$, but also that $f_{\rm HW}$ is smooth to $O(h^4)$ on AMR
boundaries. Hence if second derivatives of $f_{\rm HW}$ are computed including restricted boundary points 
in the stencil, the error will be $O(h^2)$. 

We use linear interpolation as our prolongation operator.  However, after
applying a correction from a coarse grid, we reset the values on the AMR boundaries of the 
fine grid for the points that do not exist on the coarse level with 
fourth-order interpolation using those points that do.  We found this higher order interpolation
to be beneficial as we do not relax the points on the boundary.
\bibliographystyle{h-physrev}
\bibliography{id}

\end{document}